\documentclass[10pt,journal,twocolumn,twoside]{IEEEtran}

\usepackage{cite}

\ifCLASSINFOpdf
  \usepackage[pdftex]{graphicx}
\else
  \usepackage[dvips]{graphicx}
\fi

\usepackage[cmex10]{amsmath}

\usepackage{algorithmic,algorithm,amssymb}
\usepackage[tight,footnotesize]{subfigure}
\usepackage{DefineNotations}
\usepackage{amsfonts}
\usepackage{amsbsy}
\usepackage{color,cite}

\newtheorem{theorem}{Theorem}
\newtheorem{lemma}{Lemma}

\begin{document}
\title{On the SCALE Algorithm for Multiuser Multicarrier Power Spectrum Management}
\author{Tao Wang,~\IEEEmembership{Member, IEEE}
        and Luc Vandendorpe,~\IEEEmembership{Fellow, IEEE}

\thanks{

T. Wang and L. Vandendorpe are with
ICTEAM Institute, Universit\'e Catholique de Louvain,
Louvain-la-Neuve, Belgium
(email: \{tao.wang,luc.vandendorpe\}@uclouvain.be).
The authors would like to thank the French Community for funding the project ARC SCOOP
and Fonds de la Recherche Scientifique (FNRS) in Belgium.
}}

\markboth{IEEE Trans. Signal Processing, vol. 60, no. 9, p.p. 4992 – 4998, Sep. 2012.}
{WANG AND VANDENDORPE: On the SCALE algorithm for Multiuser Multicarrier Power Spectrum Management}

\maketitle

\begin{abstract}
This paper studies the successive convex approximation for low complexity (SCALE) algorithm,
which was proposed to address the weighted sum rate (WSR) maximized dynamic power
spectrum management (DSM) problem for multiuser multicarrier systems.
To this end, we first revisit the algorithm,
and then present geometric interpretation and properties of the algorithm.
A geometric programming (GP) implementation approach is proposed and compared with the low-complexity
approach proposed previously.
In particular, an analytical method is proposed to set up the default lower-bound
constraints added by a GP solver.
Finally, numerical experiments are used to illustrate the analysis and
compare the two implementation approaches.
\end{abstract}


\IEEEpeerreviewmaketitle

\section{Introduction}

Weighted sum rate (WSR) maximized dynamic power spectrum management (DSM)
has lately been attracting much research interest for multiuser multicarrier systems.
Optimum spectrum balancing (OSB) algorithm was first proposed based on dual decomposition \cite{Cendr06}.
When there is a big number of carriers, the global optimality of this algorithm
was justified in \cite{Yu06,Luo08}, by showing that the duality gap
of the problem approaches zero asymptotically as the number of carriers goes to infinity.
A more efficient algorithm, referred to as iterative spectrum balancing (ISB),
was proposed in \cite{Cendr05,Yu06}.

Recently, a successive convex approximation for low complexity (SCALE) algorithm
was proposed based on the idea of solving convex approximations of the original problem
successively for increasingly better solutions \cite{Papan06,Papan09}.
This algorithm also proved to be very useful to address various resource allocation problems
for interference mitigation \cite{Wang11TSP-1}.
Compared with the above existing works, this paper makes the following contributions:
\begin{itemize}
\item
Novel geometric interpretation and properties are presented for the SCALE algorithm.
Most interestingly, we show that the algorithm is {\it asymptotically optimum},
i.e., it produces power allocations with WSRs approaching the maximum value 
as long as a sufficiently good initialization is used,
even though the produced power allocations are entrywise positive
while the optimum one might contain zero entries.
This property is also illustrated by numerical experiments. 

\item
A geometric programming (GP) approach is developed to implement the SCALE algorithm.
This approach reveals the fact that, each convex approximate problem for the SCALE algorithm
is actually a GP, thus a GP solver can be exploited to pursue its global optimum.
A subtlety is that default lower-bound constraints are added in the GP solver
to avoid overflow (see Appendix).
For GP based power allocation algorithms reported in the literature,
the incurred loss of optimum objective value as well as how to
set up these constraints were however not discussed \cite{Boyd07,Chiang07}.
In view of this context, these aspects are studied
for solving the DSM problem with the GP implementation of the SCALE algorithm.
The GP approach is also compared with the low-complexity implementation approach previously proposed in \cite{Papan06}.
\end{itemize}

The rest of this paper is organized as follows.
First, the system model and DSM problem are described in Section II.
In Section III, the SCALE algorithm is revisited.
Then, geometric interpretation and properties are presented in Section IV.
After that, the two implementation approaches are shown in Section V.
In Section VI, numerical experiments are given.
Finally, some conclusions and future research directions are summarized in Section VII.

{\it Notations:} A vector is denoted by a lower-case bold letter, e.g., $\bf x$,
with its $i$-th entry denoted by $[{\bf x}]_i$.
A matrix is denoted by an upper-case bold letter, e.g., $\bf X$, with
$[{\bf X}]_{ij}$ denoting the entry at the $i$-th row and $j$-th column.
${\bf X}_1\succ {\bf X}_2$ (respectively, ${\bf X}_1\succeq {\bf X}_2$)
means that ${\bf X}_1$ is entrywise strictly greater (respectively, greater) than ${\bf X}_2$. 
$e^{\bf x}$ and $e^{\bf X}$ represents the vector and matrix
which are entrywise mapped from ${\bf x}$ and ${\bf X}$ through the exponential function, respectively.
$\Der{{\bf X}}{y({\bf X})}$ stands for a matrix containing
entrywise derivative of $y({\bf X})$ with respect to ${\bf X}$.

\section{System model and DSM problem}

Consider the scenario where $K$ transmission links, each using $N$ carriers,
communicate simultaneously with cochannel interference.
Each of the transmitters and receivers is equipped with a single antenna.
Transmitter $k$ ($k=1,\cdots,K$) encodes its data and then emits them
over all carriers to receiver $k$,
with $\Pkn$ being the transmit power for carrier $n$ ($k=1,\cdots,K$, $n=1,\cdots,N$).
The channel power gain at carrier $n$ from transmitter $l$ to receiver $k$
is denoted by $\gkln$. We assume $\gkln>0$, $\forall\;l,k,n$.
It is assumed that each receiver decodes its own data by treating interference as noise,
and every coherence period in which all channels remain unchanged is sufficiently long.
There exits a spectrum management center (SMC) which first obtains $\{\gkln| \forall k,l,n\}$,
then executes a DSM algorithm,
and finally assigns the optimized power spectra to transmitters for data transmission.

To facilitate analysis, we stack all power variables into the matrix
$\Pow$ with $[\Pow]_{kn} = \Pkn$ and its $n$-th column is denoted by $\Pn$.
The signal-to-interference-plus-noise ratio (SINR) at carrier $n$ of receiver $k$
can be expressed as $\SNIRkn(\Pn) = \frac{g_{kkn}\Pkn}{\Ikn(\Pn)}$,
where $\Ikn(\Pn) = \sigma_{kn}^2 + \sum_{l=1,l\neq{k}}^K\gkln\Pln$
represents the interference-plus-noise power received at carrier $n$ for receiver $k$,
and $\sigma_{kn}^2$ is the power of additive white Gaussian noise (AWGN).
$\Gn$ is a matrix with all diagonal entries equal to zero
and $[\Gn]_{kl} = \frac{g_{kln}}{g_{kkn}}$ if $k\neq l$.
We make a mild assumption that $\forall\;n$, $\Gn$ is {\it primitive},
i.e., there exists a positive integer $i$ such that $(\Gn)^i{\succ}\Zero$
(see Theorem $8.5.2$ of \cite{Matrix-analysis}).

The WSR maximized DSM problem is
\begin{align}
\max        &\hspace{0.25cm}  g(\Pow) = \sum_{k=1}^K\sum_{n=1}^N \wk\log\big(1 + \frac{\SNIRkn(\Pn)}{\Gamma}\big)   \label{prob:original}  \\
{\rm s.t.}  &\hspace{0.25cm}  \sum_{n=1}^N\Pkn \leq \Ptotk, \forall\;k;
             \quad     \Pkn \in [0, \Pmaskkn],\forall\;k,n,                          \nonumber
\end{align}
where $\Gamma>1$, $\wk$, $\Ptotk$, and $\Pmaskkn$ represent
the SINR gap between the adopted modulation and coding scheme and the one achieving channel capacity,
the prescribed positive weight for receiver $k$'s rate,
the sum power available to transmitter $k$,
and the power spectrum mask imposed on $\Pkn$, respectively.

\section{Revisit of the SCALE algorithm}

It is difficult to find a global optimum for \eqref{prob:original}
since $g(\Pow)$ is neither convex nor concave of $\Pow$.
Initialized by a feasible $\Pow^{(1)}$, the SCALE algorithm circumvents this difficulty,
by solving convex approximations of \eqref{prob:original} successively.
To facilitate description, a superscript $m$ put to a variable indicates
that it is associated with the $m$th iteration of the algorithm hereafter ($m\geq1$).
Specifically, $\Pow^{(m+1)}$ is found as a global optimum to
an approximation of \eqref{prob:original}, i.e.,
\begin{align}
\max        &\hspace{0.25cm}  g^{(m)}(\Pow) = g(\Pow^{(m)}) + \sum_{k,n} \frac{\wk\SNIRkn^{(m)}}{\Gamma + \SNIRkn^{(m)}}
                                              \log\left(\frac{\SNIRkn(\Pn)}{\SNIRkn^{(m)}}\right)      \nonumber \\
{\rm s.t.}  &\hspace{0.25cm}  \sum_{n=1}^N\Pkn \leq \Ptotk, \forall\;k;
              \quad  \Pkn \in [0, \Pmaskkn],\forall\;k,n,                  \label{prob:SCALE-Pow}
\end{align}
where $g^{(m)}(\Pow)$ is a lower bound approximation of $g(\Pow)$ with tightness at $\Pow=\Pow^{(m)}$,
i.e., $\forall\;\Pow$, $g(\Pow) \geq g^{(m)}(\Pow)$ and $g(\Pow^{(m)}) = g^{(m)}(\Pow^{(m)})$ as proven later.
Based on this property, $g(\Pow^{(m+1)}) \geq g^{(m)}(\Pow^{(m+1)}) \geq g^{(m)}(\Pow^{(m)}) = g(\Pow^{(m)})$ follows,
ensuring that $\{g(\Pow^{(m)})|\forall\;m\}$ is an increasing sequence.
Therefore, $g(\Pow^{(m)})$ must converge as $m$ increases.

\begin{figure}[h]
\centering
     \includegraphics[width=3.5in]{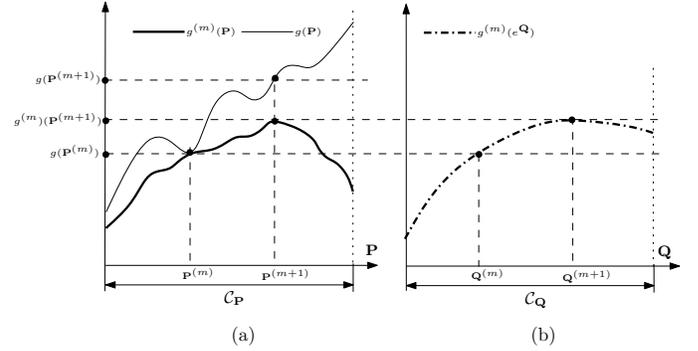}
  \caption{Illustration of the idea behind the SCALE algorithm,
           where $\regPow$ is the feasible set of \eqref{prob:original}.}
  \label{fig:SCALE-idea}
\end{figure}

Note that $g^{(m)}(\Pow)$ is not concave of $\Pow$.
Nevertheless, after replacing $\Pow$ with $\Pow = e^{\Q}$
where $\Q$ contains log-power variables ($[\Q]_{kn}=\qkn$), $g^{(m)}(e^{\Q})$ is concave of $\Q$, because
$\log(\SNIRkn(\Qn))$  is concave of $\Qn= [q_{1n}, \cdots, q_{Kn}]^T$ according to Lemma 1 of \cite{Wang11TSP-1}.
This means that after the change of variables,
\eqref{prob:SCALE-Pow} can be solved by state-of-the-art convex optimization methods.
The idea behind the SCALE algorithm is illustrated in Figure \ref{fig:SCALE-idea}.

In \cite{Papan06,Papan09}, some theoretical analysis has been made
to show that when $g(\Pow^{(m+1)}) = g(\Pow^{(m)})$, $\Pow^{(m)}$ satisfies
the Karush-Kuhn-Tucker (KKT) conditions for \eqref{prob:original}.
In the next section, we will present geometric interpretation and more properties of the SCALE algorithm.
To this end, we first revisit the derivation of $g^{(m)}(\Pow)$
as a lower bound approximation of $g(\Pow)$ with tightness at $\Pow=\Pow^{(m)}$ \cite{Papan09}.
To facilitate description, the log-SINR variable for carrier $n$ of receiver $k$ is defined as $\logSNIRkn$,
and all $\logSNIRkn$, $\forall\;k,n$ are stacked into the matrix $\logSNIR$
with $[\logSNIR]_{kn}=\logSNIRkn$.
The $\logSNIR$ corresponding to $\Q$ is denoted by $\logSNIR(\Q)$.
$[\logSNIR(\Q)]_{kn}$ is a function of $\Qn$,
and denoted by $\logSNIRkn(\Qn)$ hereafter.
Note that $\logSNIRkn(\Qn) = \log(\SNIRkn(\Qn))$ is a concave function of $\Qn$
as said earlier.
The feasible sets of $\Q$ and $\logSNIR$ are defined as $\regQ$ and
$\reglogSNIR =  \{\logSNIR(\Q) | \Q\in\regQ \}$, respectively.
It is very important to note that an analytical expression for $\reglogSNIR$
was given in \cite{Tan11JSAC}, with which it can be proven that $\reglogSNIR$ is
an unbounded convex set by using the log-convexity of the spectral radius function of a nonnegative matrix.

The WSR when expressed as $f(\logSNIR) = \sum_{k,n} \wk\log(1 + \frac{e^{\logSNIRkn}}{\Gamma})$,
is strictly convex of $\logSNIR$.
Thanks to this property, the first-order Taylor approximation of $f(\logSNIR)$
around $\logSNIR^{(m)}=\logSNIR(\Q^{(m)})$, expressed as
\begin{align}
f^{(m)}(\logSNIR) = f(\logSNIR^{(m)}) + \sum_{k,n} \Pdiff{f(\logSNIR^{(m)})}{\logSNIRkn}(\logSNIRkn - \logSNIRkn^{(m)})
\end{align}
where $\logSNIRkn^{(m)} = [\logSNIR^{(m)}]_{kn}$,
is a lower bound approximation of $f(\logSNIR)$ with tightness at $\logSNIR = \logSNIR^{(m)}$,
i.e., $\forall\;\logSNIR$, $f(\logSNIR) \geq f^{(m)}(\logSNIR)$ and $f(\logSNIR^{(m)}) = f^{(m)}(\logSNIR^{(m)})$.
Note that 
$g^{(m)}(\Pow) = f^{(m)}(\logSNIR(\Pow))$ where $\logSNIR(\Pow)$
denotes the $\logSNIR$ corresponding to $\Pow$,
indicating that $g^{(m)}(\Pow)$ is indeed a lower bound of $g(\Pow)$
with tightness at $\Pow=\Pow^{(m)}$,


\section{Analysis of the SCALE algorithm}\label{sec:GenAnalysis}

\subsection{Geometric interpretation of the algorithm over $\reglogSNIR$}

The derivation of $g^{(m)}(\Pow)$ inspires fundamental insight that,
the SCALE algorithm actually exploits the convexity of the WSR with respect to $\logSNIR$
and the entrywise concavity of $\logSNIR(\Q)$ with respect to $\Q$,
to iteratively look for increasingly better $\logSNIR\in\reglogSNIR$.
At the $m$th iteration, the first-order Taylor approximation of $f(\logSNIR)$ around $\logSNIR^{(m)}$
is used to construct $f^{(m)}(\logSNIR)$ as a lower bound approximation of $f(\logSNIR)$,
by exploiting the convexity of $f(\logSNIR)$ with respect to $\logSNIR$.
Then, a $\logSNIR$ maximizing $f^{(m)}(\logSNIR)$ over $\reglogSNIR$ is found
by solving \eqref{prob:SCALE-Pow} and assigned back to $\logSNIR^{(m+1)}$, so as to ensure that
$f(\logSNIR^{(m+1)}) \geq f^{(m)}(\logSNIR^{(m+1)}) \geq f^{(m)}(\logSNIR^{(m)}) = f(\logSNIR^{(m)})$.
{Note that this idea of formulating the WSR as a function of $\logSNIR$,
and then iteratively maximizing its first-order Taylor approximation
has also been used in \cite{Tan11JSAC,Tan11SIAM,Tan11TSP}}.

\begin{figure}[h]
  \centering
     \includegraphics[width=2in]{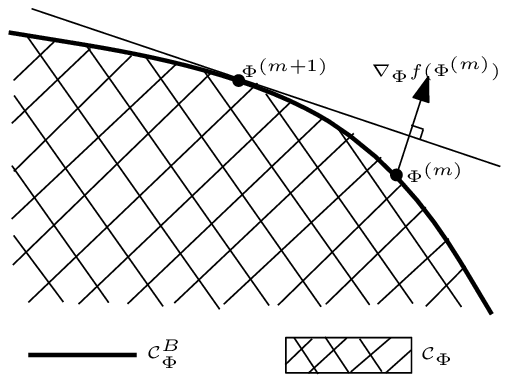}
  \caption{Illustration of looking for $\logSNIR^{(m+1)}$ over $\reglogSNIR$.}
  \label{fig:SCALE-geometric}
\end{figure}

When viewed over $\reglogSNIR$, the SCALE algorithm can be interpreted in a very simple way.
To this end, note that the $m$th iteration is to find
\begin{align}
   \logSNIR^{(m+1)} = \arg\max_{\logSNIR\in\reglogSNIR}\sum_{k,n}
                    \Pdiff{f(\logSNIR^{(m)})}{\logSNIRkn} (\logSNIRkn - \logSNIRkn^{(m)}),     \label{eq:findlogSNIRprob}
\end{align}
i.e., $\logSNIR^{(m+1)}$ is the $\logSNIR\in\reglogSNIR$ with the maximum projection
of $\logSNIR-\logSNIR^{(m)}$ toward the direction specified by $\Der{\logSNIR}f(\logSNIR^{(m)})$
as illustrated in Figure \ref{fig:SCALE-geometric}.
It can readily be shown that $\logSNIR^{(m+1)}$ must belong to
$\reglogSNIR$'s Pareto-optimal boundary (POB) denoted by $\POBreglogSNIR$,
due to the fact that $\forall\;\logSNIR^{(m)}\in\reglogSNIR$, $\Der{\logSNIR}f(\logSNIR^{(m)})\succ\Zero$.
Specifically, $\POBreglogSNIR$ consists of all $\logSNIR\in\reglogSNIR$
for which there does not exist a $\tlogSNIR\in\reglogSNIR$
satisfying $\tlogSNIR\succ\logSNIR$ \cite{Game-theory}.
Every point belonging to $\POBreglogSNIR$ represents a best tradeoff
among $\logSNIRkn$, $\forall\;k,n$ for maximizing the sum of them
weighted by certain coefficients.
Moreover, the iterative search has the following property:

\begin{lemma}\label{lemma:prop1}
If $\logSNIR^{(m+1)}\neq\logSNIR^{(m)}$, then $f(\logSNIR^{(m+1)}) > f(\logSNIR^{(m)})$.
\end{lemma}
\begin{IEEEproof}
From the strict convexity of $f(\logSNIR)$ with respect to $\logSNIR$,
$f^{(m)}(\logSNIR)$ is a strict underestimator of $f(\logSNIR)$ except at $\logSNIR = \logSNIR^{(m)}$.
Therefore, $f(\logSNIR^{(m+1)})> f^{(m)}(\logSNIR^{(m+1)}) \geq f^{(m)}(\logSNIR^{(m)}) = f(\logSNIR^{(m)})$.
\end{IEEEproof}

In a word, the SCALE algorithm iteratively searches over $\POBreglogSNIR$ to produce
$\logSNIR^{(m)}$ with strictly increasing WSR as $m$ increases\footnote{
As proposed in \cite{Papan09}, the SCALE algorithm can be generalized to maximize
the WSR of rate-adaptive (RA) users penalized by the sum power consumption of fixed-margin (FM) users.
In such a case, it can be shown that,
the SCALE algorithm can be interpreted as iteratively looking for $\logSNIR^{(m+1)}$
which maximizes the weighted sum of all $\logSNIRkn$ corresponding to the RA users
penalized by the sum power consumption of the FM users,
over a convex subset of $\reglogSNIR$ dependent on $\logSNIR^{(m)}$.}.

\subsection{Properties of the algorithm at convergence}

Note that a fixed point denoted by $\logSNIR'$ for the iterative operation
\eqref{eq:findlogSNIRprob} of the SCALE algorithm must satisfy
\begin{align}
\logSNIR' = \arg\max_{\logSNIR\in\reglogSNIR} \sum_{k,n} \Pdiff{f(\logSNIR')}{\logSNIRkn} \logSNIRkn. \label{eq:fixedpointlogSNIR}
\end{align}

Let us collect all fixed points for the SCALE algorithm in $\StaSetlogSNIR$.
It can readily be shown that $\StaSetlogSNIR$ must be part of the POB of $\reglogSNIR$,
i.e., $\StaSetlogSNIR\subset\POBreglogSNIR$.
Moreover, every $\logSNIR'\in\StaSetlogSNIR$ is a stationary point of $f(\logSNIR)$
over $\reglogSNIR$, because
\begin{align}
\forall\;\logSNIR\in\reglogSNIR, \sum_{k,n}\Pdiff{f(\logSNIR')}{\logSNIRkn} (\logSNIRkn' - \logSNIRkn)\geq 0   \label{eq:StapointlogSNIR}
\end{align}
is satisfied (see the definition of a stationary point in page $194$ of \cite{Nonlinear-opt}).
Since $\reglogSNIR$ is convex,
all local maximum of $f(\logSNIR)$ over $\reglogSNIR$ must belong to $\StaSetlogSNIR$
according to Proposition $2.1.2$ of \cite{Nonlinear-opt}.
Moreover, the following theorem can be proven:

\begin{theorem}\label{theorem:prop3}
The following claims are true:
\begin{enumerate}
\item  If $f(\logSNIR^{(m+1)}) = f(\logSNIR^{(m)})$, then $\logSNIR^{(m+1)}=\logSNIR^{(m)}\in\StaSetlogSNIR$
       holds and $\Pow^{(m+1)} = \Pow^{(m)}$ satisfies the KKT conditions of \eqref{prob:original}.
\item  Provided that
\begin{align}
   f(\logSNIR^{(1)}) > \max_{\logSNIR \in\KKTSetlogSNIR - \regoptlogSNIR} f(\logSNIR)   \label{eq:condition}
\end{align}
where $\regoptlogSNIR$ denotes the set of globally optimum $\logSNIR$ over $\reglogSNIR$,
then $\lim_{m\rightarrow+\infty}f(\logSNIR^{(m)}) = \optf$ where $\optf$ is the maximum WSR for \eqref{prob:original}.
\end{enumerate}
\end{theorem}
\begin{IEEEproof}
To prove the first claim, suppose $\logSNIR^{(m+1)} \neq \logSNIR^{(m)}$.
From Lemma \ref{lemma:prop1}, $f(\logSNIR^{(m+1)})>f(\logSNIR^{(m)})$ follows, leading to a contradiction.
Thus, $\logSNIR^{(m+1)}=\logSNIR^{(m)}\in\StaSetlogSNIR$.
Since $\logSNIR(\Pow)$ is a one-to-one mapping as shown in \cite{Tan11JSAC,Tan11SIAM},
$\Pow^{(m+1)}$ is equal to $\Pow^{(m)}$, and they satisfy the KKT conditions of \eqref{prob:original} as proved in \cite{Papan06}.

To prove the second claim, two cases are examined.
In the first case, an $m$ satisfying $f(\logSNIR^{(m)}) = f(\logSNIR^{(m+1)})$ exists.
Suppose $\logSNIR^{(m)}\notin\regoptlogSNIR$, then $\logSNIR^{(m)}\in \KKTSetlogSNIR-\regoptlogSNIR$ follows,
meaning that $f(\logSNIR^{(1)})>f(\logSNIR^{(m)})$.
This is a contradiction with the fact that $f(\logSNIR^{(m)})$ is strictly increasing with $m$,
thus $\logSNIR^{(m)}\in\regoptlogSNIR$ and $\lim_{m\rightarrow+\infty}f(\logSNIR^{(m)}) = \optf$ holds for the first case.
In the second case, $f(\logSNIR^{(m)})$ strictly increases endlessly
as illustrated by a numerical example in Section VI-A.
According to the monotone convergence theorem,
$f(\logSNIR^{(m)})$ approaches the supremum of $\{f(\logSNIR)|\logSNIR\in\POBreglogSNIR\}$, i.e.,
$\lim_{m\rightarrow+\infty}f(\logSNIR^{(m)})= \sup_{\logSNIR\in\POBreglogSNIR}f(\logSNIR)= \optf$ \cite{Monotonic-converge}.
Therefore, the second claim is true.
\end{IEEEproof}

The first claim indicates that when $f(\logSNIR^{(m+1)}) = f(\logSNIR^{(m)})$,
the SCALE algorithm reaches a fixed point in $\StaSetlogSNIR$,
and $\Pow^{(m)}$ (respectively, $\logSNIR^{(m)}$) will remain there for all following iterations,
i.e., it never happens that $f(\Pow^{(m)})$ remains fixed while
$\Pow^{(m)}$ keeps oscillating among different values as $m$ increases.
Nevertheless,  it may happen that $f(\Pow^{(m)})$ increases strictly and endlessly
as illustrated by a numerical example in Section VI-A.
Therefore, the SCALE algorithm can be terminated when either prescribed
$M$ iterations are already executed or $\Pow^{(m+1)} = \Pow^{(m)}$ is satisfied.

According to the second claim of Theorem \ref{theorem:prop3},
the SCALE algorithm is asymptotically optimum, i.e., it produces $\Pow^{(m)}$ with WSR approaching $\optf$
as long as $f(\Pow^{(1)})$ is above the threshold value on the right-hand side of \eqref{eq:condition},
even though $\forall\;m$, $\Pow^{(m)}$ is entrywise positive
while the optimum $\Pow$ for \eqref{prob:original} might contain zero entries\footnote{
This happens under certain conditions shown in \cite{Luo07}.
However, it still remains an open and challenging problem to decide which users and carriers should be shut down.}.
This will be illustrated by a numerical example in Section VI-A.
Note that the above condition is generally true to guarantee global optimality.
It was also pointed out in \cite{Tan11JSAC,Tan11SIAM,Tan11TSP} that for the single carrier case,
the SCALE algorithm always converges to a global optimum
when the optimum $\logSNIR$ and $\{\wk|\forall\;k\}$ satisfy some special conditions.

\section{Approaches to implement the SCALE algorithm}

The key to implementing the SCALE algorithm is to solve the $m$th iteration problem:
\begin{align}
\max_{\Q}  &\hspace{0.25cm}  \sum_{k,n} \amkn\big(\qkn -\log \big( \Nvarkn + \sum_{l\neq{k}} \gkln e^{\qln}) \big)  \label{prob:SCALE-Q2} \\
{\rm s.t.} &\hspace{0.25cm}  \sum_{n=1}^N e^{\qkn} \leq \Ptotk, \forall\;k;
           \quad  e^{\qkn} \leq \Pmaskkn,\forall\;k,n,  \nonumber
\end{align}
where $\amkn = \frac{\wk\SNIRkn^{(m)}}{\Gamma + \SNIRkn^{(m)}}
= \frac{\wk\gkkn\Pkn^{(m)}}{\gkkn\Pkn^{(m)} + \Gamma\cdot\Ikn(\Pow^{(m)})}$.
In the following, we first present the low-complexity approach proposed in \cite{Papan06},
and then the GP approach to implement the SCALE algorithm. Finally, the two approaches are compared.

\subsection{The low-complexity approach}

\begin{algorithm}
\caption{The low-complexity implementation approach}\label{Alg:dual}
\begin{algorithmic}[1]
\STATE  $m=0$; $\forall\;k,n$, $\amkn=\wk$, $\Pkn=0$;
\REPEAT
        \REPEAT
                \STATE  $\forall\;k$, $\mu_k=0$ if $\sum_n \newPkn(\Pow,0) \leq \Ptotk$,
                        otherwise search the $b_k>0$ satisfying $\sum_n \newPkn(\Pow,b_k) = \Ptotk$ with the bisection method,
                        and $\mu_k = b_k$;
                \STATE  $\forall\;k,n$, update $\Pkn$ with $\newPkn(\Pow,\mu_k)$;
        \UNTIL{$\Pow$ converges or $L$ iterations have been executed}
        \STATE  $m=m+1$; $\Pow^{(m)} = \Pow$; compute $\amkn$, $\forall\;k,n$;
\UNTIL{$m = M$ or $\Pow^{(m)} = \Pow^{(m-1)}$}
\STATE output $\Pow^m$ as a solution to \eqref{prob:original}.
\end{algorithmic}
\end{algorithm}

A low-complexity approach was proposed in \cite{Papan06} to solve \eqref{prob:SCALE-Q2},
with KKT conditions based fixed-point equations
and the bisection method for updating $\Pow$ and Lagrange multipliers, respectively.
To facilitate description, the Lagrange multiplier for transmitter $k$'s
sum power constraint is denoted as $\mu_k$.
The KKT conditions require the $\Pow$ corresponding to the optimum $\Q$
and the optimum $\mu_k$, $\forall\;k$ to satisfy:
\begin{align}
\forall\;k,n,\; \Pkn = \newPkn(\Pow,\mu_k), {\;\rm and\;}
\forall\;k,\; \mu_k \left(\sum_n \Pkn - \Ptotk\right) = 0,  \nonumber 
\end{align}
where
$\newPkn(\Pow,\mu_k)
= \left[ \frac{\amkn}{\mu_k + \sum_{l\neq k}(\amln\glkn/\Iln(\Pow))} \right]_{0}^{\Pmaskkn}$
and $[x]_{y}^{z} = \max\{y,\min\{x,z\}\}$.
The low-complexity approach is summarized in Algorithm \ref{Alg:dual}.

\subsection{The GP approach}

In fact, it can be shown that \eqref{prob:SCALE-Q2} is equivalent to
\begin{align}
\min_{\Q}  &\hspace{0.25cm}  \log\big(e^{\sum_{k,n} \amkn(\tkn - \qkn)}\big)   \nonumber \\
{\rm s.t.} &\hspace{0.25cm}  \log\big(\sum_{n} \frac{1}{\Ptotk}e^{\qkn}\big) \leq 0,                                  \forall\;k,  \nonumber  \\
           &\hspace{0.25cm}  \log\big(\sum_{l\neq k}\gkln e^{\qln - \tkn} + \sigma_{kn}^2 e^{-\tkn}\big) \leq 0,    \forall\;k,n,    \label{prob:SCALE-Q2-GP}  \\
           &\hspace{0.25cm}  \log(\frac{1}{\Pmaskkn} e^{\qkn})\leq0,                                                    \forall\;k,n.  \nonumber
\end{align}
where $\{\tkn|\forall\;k,n\}$ is a set of extra variables introduced to guarantee equivalence.
The GP approach to implement the SCALE algorithm simply uses a GP solver,
to solve \eqref{prob:SCALE-Q2-GP} for $\Q^{m+1}$ (see Appendix).

A subtlety deserving special attention is that,
an extra lower-bound constraint on every $\qkn$, expressed as $\qkn \geq \log(\xi)$
where $\xi$ can be preassigned,
is added by default in the GP solver to avoid overflow\footnote{
As for $\tkn,\forall\;k,n$, the idle constraint $\tkn\geq \log(\sigma_{kn}^2)$,
i.e., it is always relaxed since the second constraint in \eqref{prob:SCALE-Q2-GP} should be satisfied,
can be preassigned to the GP solver. }.
Denote the feasible set of $\logSNIR$  by $\reglogSNIR(\xi)$
after the extra constraints $\Pkn \geq \xi$, $\forall\;k,n$ are added. 
It can be seen that after adding the extra constraints, the GP approach iteratively searches
over $\POBreglogSNIR(\xi)$ representing the POB of $\reglogSNIR(\xi)$ 
to produce $\Pow^{(m)}$ with strictly increasing WSR.

The GP approach actually implements the SCALE algorithm
to solve \eqref{prob:original} with the extra constraints.
It is interesting to evaluate the incurred loss of the maximum WSR
for \eqref{prob:original} due to the extra constraints.
Obviously, $\optf(\xi) \leq \optf$ where $\optf(\xi)$ is the global optimum for
\eqref{prob:original} with the extra constraints.
Note that \eqref{prob:original} might have multiple globally optimum solutions,
one of which is denoted by $\optPow$ hereafter.
If there exists a $\optPow$ satisfying $\optPow\succeq\xi$, $\optf(\xi) = \optf$ follows.
Otherwise, $\optf(\xi) < \optf$, meaning that a loss of the optimum WSR is incurred by the extra lower bounds.
In practice, it is rarely known a priori if there always exists a $\optPow$ satisfying $\optPow\succeq\xi$.
To evaluate the worst-case loss of the optimum WSR, we present the following theorem:

\begin{theorem}\label{theorem:GP}
Suppose $0\leq\xi\ll\frac{\Ptotk}{N(N+1)}$, then
\begin{align}
\optf\geq \optf(\xi) \geq \optf - 2\xi N K^2 \max_{k,l,n}{\frac{w_l\glkn}{\sigma^2_{ln}}} \label{eq:WSRloss-GP}
\end{align}
\end{theorem}
\begin{IEEEproof}
To prove the claim, two cases are examined.
In the first case, there exists a $\optPow\succeq\xi$,
thus \eqref{eq:WSRloss-GP} follows since $\optf(\xi) = \optf$.
In the second case, there exists at least an entry smaller than $\xi$ for any $\optPow$,
and we will prove the validity of \eqref{eq:WSRloss-GP} as follows.

Let's first choose a $\optPow$, from which we will construct a $\Pow''$
very close to it and feasible for \eqref{prob:original} with the extra constraints in two steps.
In the first step, all entries in $\optPow$ smaller than $\xi$ are raised to be $\xi$.
The resulting power allocation is denoted by $\Pow'$.
Clearly, the total increase of every transmitter's sum power is not higher than $N\xi$.
In the second step, all rows of $\Pow'$ are examined.
For the $k$th row, no entries is updated if $\sum_n{[\Pow']_{kn}}\leq\Ptotk$.
Otherwise, only the maximum entry, i.e., $[\Pow']_{kn_k}$ with $n_k = \arg\max_{n}[\Pow']_{kn}$,
which must satisfy $[\Pow']_{kn_k} \geq \frac{\Ptotk}{N} \gg (N+1)\xi$, is reduced by $N\xi$.
The finally produced power can be taken as $\Pow''$.

It can be shown that $\forall\;k,n$,
$\frac{\partial g(\optPow)}{\partial\Pkn} = \frac{\partial C_n(\optPow)}{\partial\Pkn} - \frac{\partial D_n(\optPow)}{\Pkn}$
where $C_n(\Pow) = \sum_{l} w_l\log(\glln\Pln + \Gamma\Iln(\Pow))$ and
$D_n(\Pow) = \sum_{l} w_l\log(\Gamma\Iln(\Pow))$.
Therefore,
\begin{align}
&\frac{\partial g(\optPow)}{\partial\Pkn} \geq -\frac{\partial D_n(\optPow)}{\Pkn}
                                          = -\sum_{l\neq k}\frac{w_l\glkn}{\Iln(\optPow)}
                                          \geq -K\max_{k,l,n}{\frac{w_l\glkn}{\sigma^2_{ln}}},  \nonumber\\
&\frac{\partial g(\optPow)}{\partial\Pkn} \leq \frac{\partial C_n(\optPow)}{\partial\Pkn}
                                          \leq K\max_{k,l,n}{\frac{w_l\glkn}{\sigma^2_{ln}}}. \nonumber
\end{align}

Obviously, $\optf \geq \optf(\xi) \geq g(\Pow'')$.
Since $\xi$ is very small, $g(\Pow'')$ can be computed according to
the first-order Taylor approximation around $\optPow$ as
\begin{align}
g(\Pow'') &\approx \optf + \sum_{k\in \mathcal{S}_k}\left(
                -\delta\frac{\partial f(\optPow)}{\partial p_{kn_k}} +
                \sum_{n\in \mathcal{S}_{kn}}\xi\frac{\partial f(\optPow)}{\partial \Pkn} \right),   \nonumber
\end{align}
where $\delta$ is equal to either $0$ or $N\xi$,
$k$ belongs to $\mathcal{S}_k$ if the entries in $\optPow$ corresponding to transmitter $k$
are modified to get $\Pow''$,
and $\mathcal{S}_{kn}$ is the set of carrier numbers $n$ satisfying $[\optPow]_{kn}<\xi$.
Therefore,
\begin{align}
g(\Pow'') &\geq \optf + \sum_{k\in \mathcal{S}_k} \left(
                - \delta K\max_{k,l,n}{\frac{w_l\glkn}{\sigma^2_{ln}}}
                -\xi NK\max_{k,l,n}{\frac{w_l\glkn}{\sigma^2_{ln}}} \right)        \nonumber\\
          &\geq \optf - 2\xi N K^2 \max_{k,l,n}{\frac{w_l\glkn}{\sigma^2_{ln}}},  \nonumber
\end{align}
meaning that the claim holds for the second case as well.
\end{IEEEproof}

According to the above theorem, $\xi$ should satisfy
$\xi \leq \epsilon\left(2 N K^2 \max_{k,l,n}{\frac{w_l\glkn}{\sigma^2_{ln}}}\right)^{-1}$
to ensure that $\optf(\xi)\geq \optf-\epsilon$,
where $\epsilon >0$ is prescribed and represent the maximum tolerable loss of the optimum WSR.
Algorithm \ref{Alg:GP} summarizes the GP approach to implement the SCALE algorithm,
where a $\xi$ satisfying the above condition is used to configure the GP solver.

\begin{algorithm}
\caption{The GP implementation approach}\label{Alg:GP}
\begin{algorithmic}[1]
\STATE  $m=0$; $\forall\;k,n$, $\amkn = w_k$;
\REPEAT
        \STATE solve \eqref{prob:SCALE-Q2-GP} for $\Q^{(m+1)}$ by a GP solver configured
               with a $\xi \leq \epsilon\left(2 N K^2 \max_{k,l,n}{\frac{w_l\glkn}{\sigma^2_{ln}}}\right)^{-1}$;
        \STATE $\Pow^{(m+1)}=e^{\Q^{(m+1)}}$; $m = m+1$; compute $\amkn$, $\forall\;k,n$;
\UNTIL{$m=M$ or $\Pow^{(m)} = \Pow^{(m-1)}$}
\STATE output $\Pow^{(m)}$ as a solution to \eqref{prob:original}.
\end{algorithmic}
\end{algorithm}

\subsection{Comparison of the two approaches}

As said earlier, the GP approach iteratively searches over $\POBreglogSNIR(\xi)$
to produce $\Pow^{(m)}$ with strictly increasing WSRs, i.e., Lemma \ref{lemma:prop1} still holds.
Therefore, the GP approach has guaranteed convergence as $m$ increases.
Note that {\it $\reglogSNIR(\xi)$ might be nonconvex} as illustrated later.
Theorem \ref{theorem:prop3} also remains true when $\StaSetlogSNIR$ and $\optf$
are replaced by $\StaSetlogSNIR(\xi)$ and $\optf(\xi)$, respectively.
This means that as long as a sufficiently good initialization is used,
the GP approach still produces $\Pow^{(m)}$ with WSR approaching $\optf(\xi)$ asymptotically.

The low-complexity approach uses a heuristic rule
to pursue the combination satisfying the KKT conditions of \eqref{prob:SCALE-Q2}.
Once converged, the inner iterations must output a global optimum for \eqref{prob:SCALE-Q2} as $\Pow^{(m+1)}$.
If the convergence of inner iterations could always be achieved,
the low-complexity approach would become a faithful implementation of the SCALE algorithm.
However, it is unclear if the inner iterations always converge as $L$ increases,
since a theoretical proof is difficult and has not been available yet.
Nevertheless, numerical experiments show that the convergence is always observed
when $L$ is sufficiently large for practical channel realizations \cite{Papan06}.
In practice, it is very attractive to implement the SCALE algorithm
with the low-complexity approach using a very small $L$.
In such a case, the inner iterations might often not converge,
and the low-complexity approach might have very interesting behavior, e.g.,
{\it the produced $f(\Pow^{(m+1)})$ might be either smaller,
or even greater than the WSR of the optimum for \eqref{prob:SCALE-Q2},
as will be illustrated later.}

Note that the GP approach with a very small $\xi$
can be regarded as a close approximation of the faithful implementation of the SCALE algorithm.
Therefore, the GP approach can be used as a benchmark to evaluate the low-complexity approach.
In Section VI, we will further use numerical experiments to compare
the GP approach and the low-complexity approach,
and show the behavior of the low-complexity approach when $L$ varies.

\section{Numerical experiments}

\subsection{A simple numerical example to illustrate analysis}

\begin{table}[!b]
\caption{Three sets of channel and weight parameters.}\label{tab:paras}
\center
\begin{tabular}{c|cccccc}
               & $g_{11}$ & $g_{22}$ & $g_{12}$ & $g_{21}$ & $w_1$ & $w_2$ \\
  \hline\hline
  the first set  &1.0      &  1.0     & 0.8    & 0.4    & 3.0 & 1.0 \\
  \hline
  the second set  & 1.0      &  1.0     & 1.8    & 0.4    & 1.8 & 1.0 \\
  \hline
\end{tabular}
\end{table}

We first consider a simple scenario with $K=2$ and $N=1$
because a faithful implementation of the SCALE algorithm can readily be made in this scenario.
The parameters are set as $\Gamma=0$ dB, $\forall\;k,n$, $\Ptotk=\Pmaskkn=\gkkn=\sigma^2_{kn}=1$.
Since a single carrier exists, the carrier-number subscript of
every variable is omitted for simplicity hereafter.
The analytical expression $\reglogSNIR$ and $\reglogSNIR(\xi)$ can readily be derived and
thus not shown here due to space limitation.

\begin{figure}
  \centering
  \subfigure[]{
     \includegraphics[width=3in]{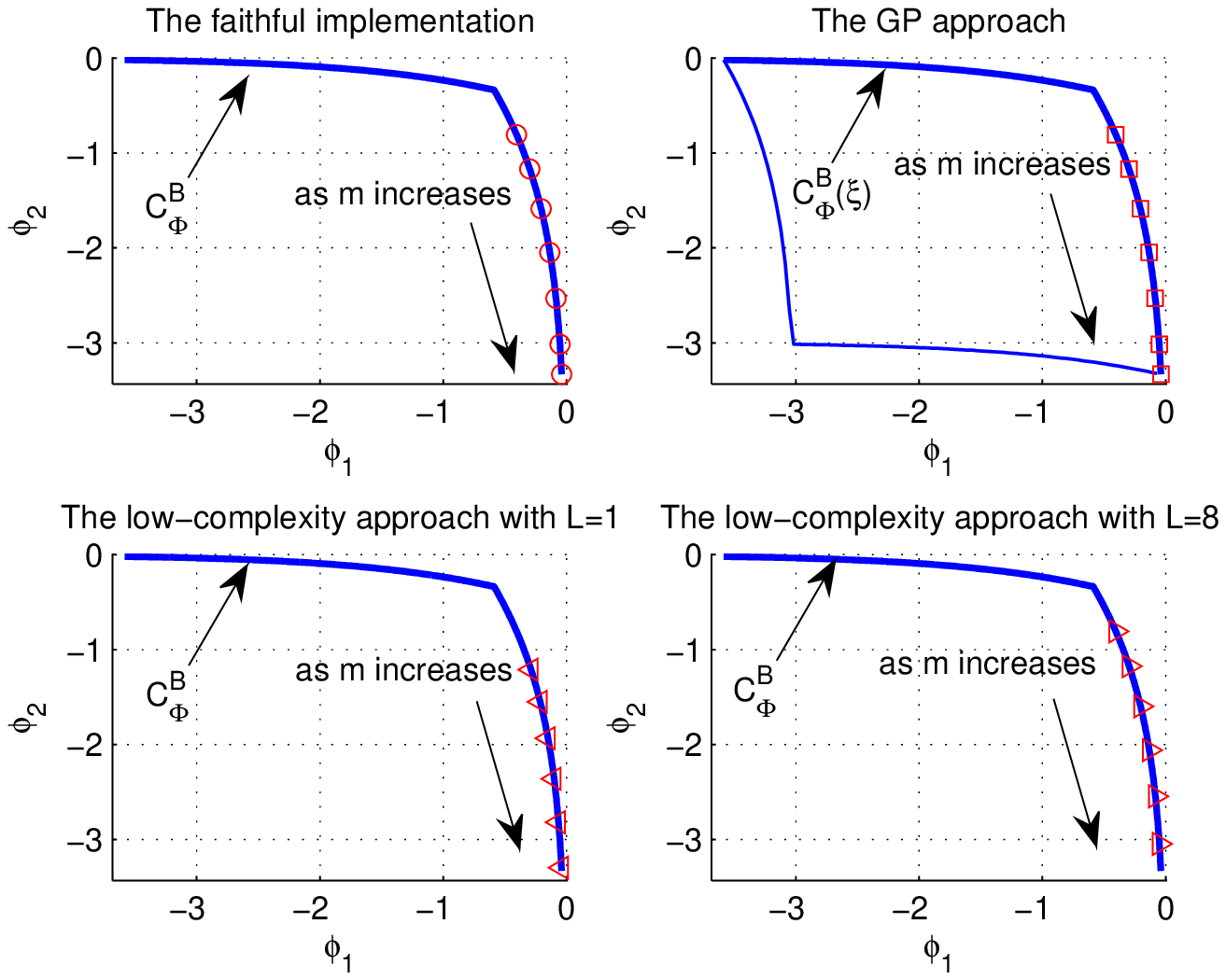}}
  \subfigure[]{
     \includegraphics[width=3in]{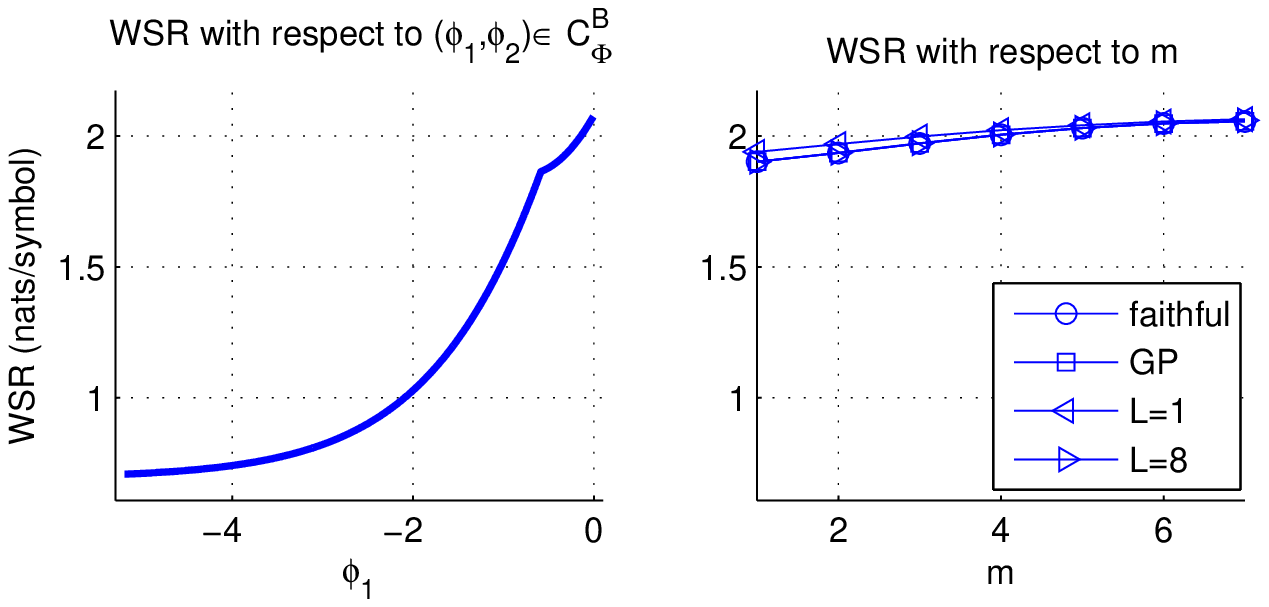}}
  \caption{Visualization of the iterative searches for the approaches to implement the SCALE algorithm,
           when the second set of channel and weight parameters is used. }
  \label{fig:set2}
\end{figure}

Figure \ref{fig:set2} shows the results when the first set of channel and weight parameters
in Table \ref{tab:paras} is used.
For the faithful and low-complexity approaches illustrated in Figure \ref{fig:set2}.a,
$\POBreglogSNIR$ is plot as the line asymptotically extending to $(-\infty,0)$ and $(0,-\infty)$, respectively,
and $\reglogSNIR$ is the unbounded region below that line.
For the GP approach, $\xi=0.05$ is chosen, and the POB and the lower-boundary for $\reglogSNIR(\xi)$ are plot,
with $\reglogSNIR(\xi)$ being the region enclosed inside.
Obviously, $\reglogSNIR(\xi)\subseteq\reglogSNIR$ and $\POBreglogSNIR(\xi)\subseteq\POBreglogSNIR$, respectively.
$\reglogSNIR$ is convex, which illustrates the validity of the analysis in \cite{Tan11JSAC,Tan11SIAM,Tan11TSP}.
Most interestingly, $\reglogSNIR(\xi)$ is nonconvex for this parameter set.
The globally optimum $\logSNIR$ in $\reglogSNIR$ and $\reglogSNIR(\xi)$
is at $(\phi_1=0,\phi_2=-\infty)$ and $(\phi_1=-0.04,\phi_2=-3.33)$,
and corresponds to $\optf = 2.08$ and $\optf(\xi)=2.06$, respectively.
Obviously, they satisfy \eqref{eq:WSRloss-GP}, thus illustrating the validity of Theorem \ref{theorem:GP}.
The faithful implementation and the low-complexity approach with $L=8$ produce the same sequence
of $\logSNIR^m$ approaching the globally optimum $\logSNIR$ asymptotically as $m$ increases
(the iterations proceed endlessly, and only the first $7$ iterations are shown here for clarity).
{\it This illustrates the asymptotic optimality of the SCALE algorithm
even when the optimum $\Pow$ contains zero entries, as discussed in Section IV. }
This also suggests that the inner iterations for the low-complexity approach with $L=8$
always converge to an optimum solution for \eqref{prob:SCALE-Q2}, $\forall\;m$.
The GP approach converges to the globally optimum $\logSNIR$ in $\reglogSNIR(\xi)$ after $7$ iterations.
On the other hand, the low-complexity approach with $L=1$ produces a different sequence of solutions,
suggesting its inner iterations do not always converge to the global optimum for \eqref{prob:SCALE-Q2}.
Nevertheless, the produced $\logSNIR^{(m)}$ still approaches the globally optimum $\logSNIR$ asymptotically.

\begin{figure}
  \centering
  \subfigure[]{
     \includegraphics[width=3in]{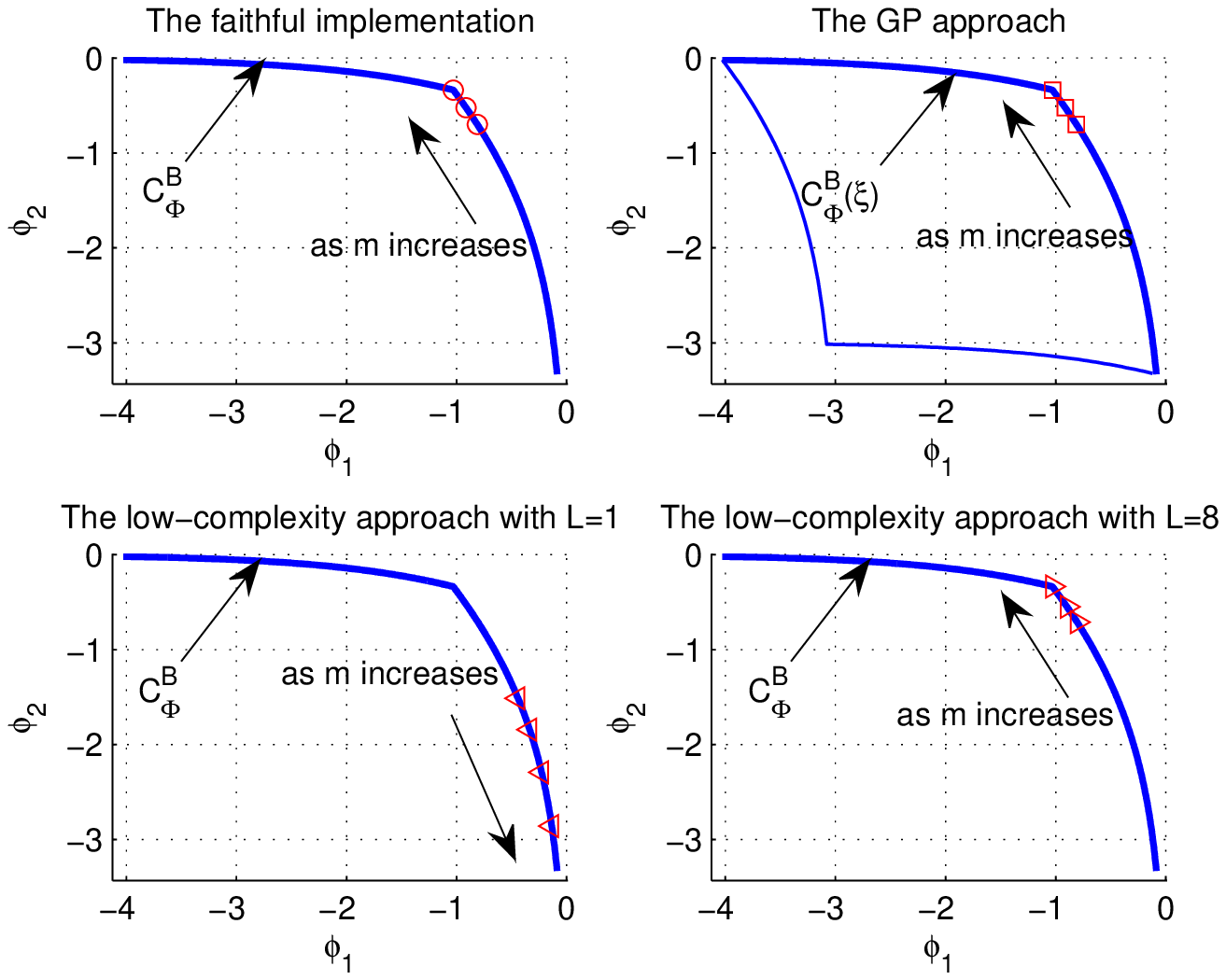}}
  \subfigure[]{
     \includegraphics[width=3in]{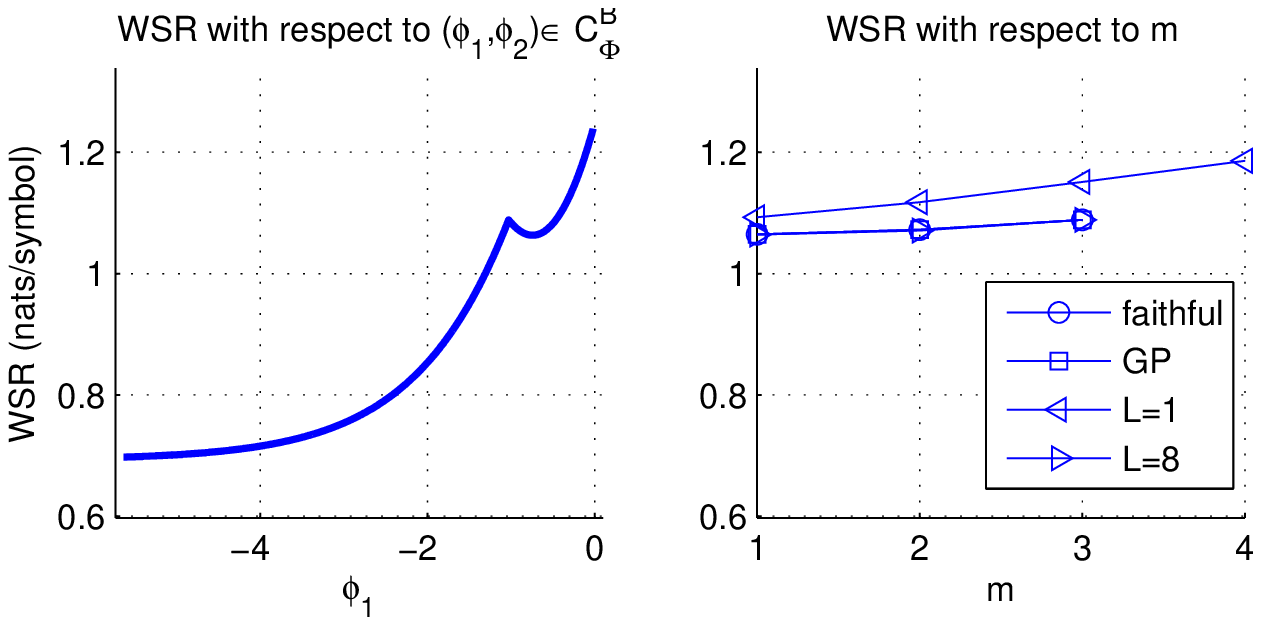}}
  \caption{Visualization of the iterative searches for the approaches to implement the SCALE algorithm,
           when the third set of channel and weight parameters is used. }
  \label{fig:set3}
\end{figure}

For the second parameter set in Table \ref{tab:paras}, the results are shown in Figure \ref{fig:set3}.
For this parameter set, the globally optimum $\logSNIR$ in $\reglogSNIR$ and $\reglogSNIR(\xi)$
is at $(\phi_1=0,\phi_2=-\infty)$ and $(\phi_1=-0.12,\phi_2=-3.32)$,
and corresponds to $\optf = 1.25$ and $\optf(\xi) = 1.17$, respectively.
They still satisfy \eqref{eq:WSRloss-GP} and illustrate the validity of Theorem \ref{theorem:GP}.
Similar phenomena can be observed as for the first parameter set, except for those explained as follows.
The faithful, GP and low-complexity approaches with $L=8$ all produce the same sequence of $\logSNIR^{(m)}$,
and converge to a fixed point $(\phi_1=-1.03,\phi_2=-0.34)$ which is locally optimum after $3$ iterations.
In particular, the $f(\logSNIR^{(1)})$ produced by all the approaches does not satisfy condition \eqref{eq:condition},
indicating such a convergence to a local optimum is indeed possible.
Moreover, it is very interesting to see that for the low-complexity approach with $L=1$,
the produced $\logSNIR^{(1)}$ has a higher WSR than that for the other approaches,
and the $\logSNIR^{(m)}$ approaches the globally optimum $\logSNIR$ asymptotically as $m$ increases
(only the first $4$ iterations are shown here).
{\it This suggests that the low-complexity approach using a small $L$,
even without convergence of its inner iterations, might lead to a better solution than the GP approach. }

\subsection{Numerical experiments for a realistic scenario}

We have also conducted numerical experiments for a realistic scenario with $K = 4$ and $N=128$.
The $k$th receiver is located at the coordinate $(x=k,y=10)$,
whereas the $k$th transmitter is at $(x=k,y=5)$ and $(x=k,y=0)$
if $k=1,2$ and $k=3,4$, respectively. These coordinates are in the unit of meter.
The parameters are set as $\Gamma=0$ dB, $w_1=w_2=1$, $w_3=w_4=2$,
$\forall\;k,n$, $\sigma^2_{kn}= -30$ dBm, $\Pmaskkn=\Ptotk$.
The channel for every link is generated with the channel model explained in \cite{Wang11TSP-2,Wang11JSAC}.
Note that the transmitted power is attenuated by $30$ dB
in average when received at a distance of $10$ meter apart.
To evaluate the effectiveness of the implementation approaches for the SCALE algorithm,
the ISB algorithm proposed in \cite{Yu06} was also implemented
(note that a grid search of $100$ points was used to optimize every power variable to ensure
a good performance).
We conducted the following experiments with Matlab v7.1 on a laptop equipped
with an Intel Duo CPU of 2.53 GHz and a memory of 3 GBytes.

We have generated $100$ random realizations of all channels.
When $\forall\;k$, $\Ptotk=50$ dBm, we have implemented for every realization
the ISB algorithm, the GP and low-complexity approaches with different $L$.
The GP solver gpcvx was used to solve \eqref{prob:SCALE-Q2-GP} for the GP approach,
and $\xi = 10^{-10}$ is assigned \cite{gpcvx}.
During the simulation, we find that for every channel realization,
the maximum $g_{lkn}$, $\forall\;k,l,n$ is smaller than $10^{-3}$.
According to Theorem \ref{theorem:GP}, $\optf(\xi)\geq \optf - 8.19\times10^{-7}$ follows,
meaning that the worst-case loss of the optimum WSR is negligible.

The average time spent by the ISB algorithm is around $600$ seconds.
The average time for the GP approach increases proportionally with $M$ and it is around $100$ seconds when $M=8$,
and the one for the low-complexity approach increases proportionally with $L*M$,
and it is around $1.5$ seconds when $M=8$ and $L=16$.
The GP approach is much faster than the ISB algorithm,
because the interior-point method (IPM) used is more efficient than the dual method and exhaustive search used by the ISB algorithm.
The low-complexity approach is much faster than the GP approach,
indicating that the heuristic update rule used for the inner iteration of the low-complexity approach
leads to a much faster speed than the IPM used by the GP solver.

\begin{figure}\centering
     \includegraphics[width=3in]{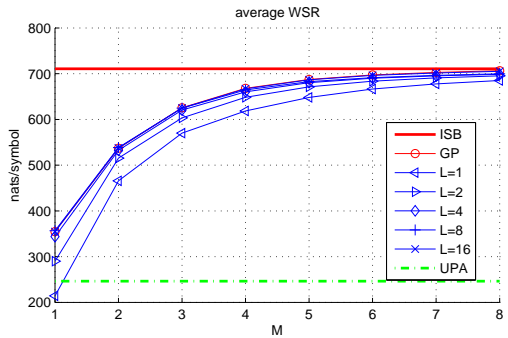}
  \caption{The average WSRs produced by every implementation approach for the SCALE algorithm,
            and for using the UPA and ISB algorithm, respectively. }
  \label{fig:avg}
\end{figure}

Figure \ref{fig:avg} shows the average WSR of the solutions produced by every
implementation approach for the SCALE algorithm when $M\leq 8$.
The average WSRs corresponding to using the uniform power allocation (UPA)
and the ISB algorithm, respectively, are also shown.
It can be seen that the SCALE algorithm implemented with every approach
leads to a much greater average WSR than the UPA when $M\geq 2$.
The average WSR for the GP approach increases and becomes close to that for the ISB approach
as $M$ increases, which illustrates the effectiveness of the SCALE algorithm.
For the low-complexity approach with a fixed $L$, the average WSR increases as $M$ increases.
For the low-complexity approach with a fixed $M$,
the average WSR increases as $L$ increases, and is close to that for the GP approach when $L\geq 4$.
Most interestingly, the low-complexity approach using $L=1$ and $M=8$
is a good option to implement the SCALE algorithm in practice,
since it has a fast speed and its average WSR performance is close to that for the GP approach.

\section{Conclusion and future work}
We have presented the geometric interpretation and properties of the SCALE algorithm.
A GP approach has also been developed to implement the algorithm,
and an analytical method to set up the lower-bound constraints added by a GP solver.
Numerical experiments have been shown to illustrate the analysis
and compare the GP approach and the low-complexity approach proposed previously.

In future, the following aspects can be further investigated.
First, a theoretical study can be made on the convergence of
the inner iterations for the low-complexity approach,
as well as the behavior of the low-complexity approach using a small inner iteration number.
Second, we can study how to generalize the SCALE algorithm
for multiple-input-multiple-output systems,
and compare it with other algorithms \cite{WMMSE-1,WMMSE-2}.
Third, it is important to note that the SCALE algorithm iteratively
produces increasingly better $\logSNIR$ through solving \eqref{prob:SCALE-Pow} for
the globally optimum $\Q\in\regQ$ and then transforming it back to $\logSNIR$.
Another possible way is to solve \eqref{eq:findlogSNIRprob} directly for
the globally optimum $\logSNIR\in\reglogSNIR$, e.g., by using an analytical formulation of $\reglogSNIR$.
Works along this direction have been done in \cite{Tan11JSAC,Tan11SIAM,Tan11TSP}
for the single-carrier case. It is interesting to check how to extend those studies
for the multicarrier case.

\section*{Acknowledgement}

The authors would like to thank Prof. Min Dong and the anonymous reviewers
for their precious comments and suggestions which significantly improve this work.



\appendix

\section{A brief introduction to geometric programming}

A standard-form GP problem is expressed as
\begin{align}
\min_{\xvec}  &\hspace{0.25cm}   f_0(\xvec)                          \nonumber \\
{\rm s.t.}    &\hspace{0.25cm}   f_j(\xvec) \leq 1, j = 1,..., J,    \label{prob:GP-standard}\\
              &\hspace{0.25cm}   x_i\geq 0, i = 1,..., I,            \nonumber
\end{align}
where $\forall\;j=0,1,\cdots$, $f_j(\xvec)$ is a posynomial of $\xvec$ with $[\xvec]_i = x_i$.
Specifically, an example posynomial of $\xvec$ is expressed as
$p(\xvec) = \sum_{a=1}^{A} u_{a}(\xvec)$ with $u_{a}(\xvec) = \beta_{a}\prod_{i=1}^I x_i^{\alpha_{ai}}$
where $\beta_{a}>0$ and $\alpha_{ai}$ ($i=1,\cdots,I$) are real constants.
It is very important to note that $s(\yvec) = \log\big(p(e^{\yvec})\big)$ is convex of $\yvec$.
Therefore, although a standard-form GP problem in \eqref{prob:GP-standard} is nonconvex,
it can be converted by making the logarithmic transformation from $\xvec$ to $\yvec$
satisfying $\xvec = e^{\yvec}$ to its equivalent convex form
\begin{align}
\min_{\yvec}  &\hspace{0.25cm}   g_0(\yvec) = \log\big(f_0(e^{\yvec})\big)                        \label{prob:GP-convex}\\
{\rm s.t.}    &\hspace{0.25cm}   g_j(\yvec) = \log\big(f_j(e^{\yvec})\big) \leq 0, j = 1,..., J,  \nonumber
\end{align}
which is then solved by a GP solver, e.g., gpcvx or MOSEK based on state-of-the-art IPM.

An important subtlety should be noted.
When all globally optimum $\xvec$ for \eqref{prob:GP-standard} contains zero entries,
the corresponding optimum $\yvec$ for \eqref{prob:GP-convex} contains entries equal to $-\infty$.
In such a case, the IPM iteratively outputs $\yvec$ with $g_0(\yvec)$ asymptotically approaching
the optimum objective value for \eqref{prob:GP-convex},
which leads to an overflow in the processor running the GP solver that implements the IPM.
It is rarely known a priori if the optimum $\xvec$ for \eqref{prob:GP-standard} contains zero entries.
To avoid overflow, the GP solver by default adds entrywise lower-bound constraints on $\xvec$,
or equivalently on $\yvec$ before solving \eqref{prob:GP-convex} (see page 3 of \cite{gpcvx}).
These constraints should be set to ensure that the optimum $\xvec$ for \eqref{prob:GP-standard}
with the extra constraints corresponds to an objective value within
a prescribed small tolerance around the original optimum objective for \eqref{prob:GP-standard}.





\ifCLASSOPTIONcaptionsoff
  \newpage
\fi

\bibliographystyle{IEEEtran}
\bibliography{Multicell-Reference}

\end{document}